\documentclass[aps,prb,twocolumn,amsmath,amssymb,superscriptaddress]{revtex4-1}
\usepackage{graphicx}
\usepackage{dcolumn} 
\usepackage{graphics}
\usepackage{bbm}
\usepackage{amsmath,amsfonts,amssymb}
\usepackage{wrapfig}
\usepackage[pdftex]{color}

\begin{document}
\title{Stimulated Phonon Emission in a Driven Double Quantum Dot}
\author{J. I. Colless$^*$}
\affiliation{ARC Centre of Excellence for Engineered Quantum Systems, School of Physics, The University of Sydney, Sydney, NSW 2006, Australia.} 
\author{X. G. Croot$^*$}
\affiliation{ARC Centre of Excellence for Engineered Quantum Systems, School of Physics, The University of Sydney, Sydney, NSW 2006, Australia.} 
\author{T. M. Stace}
\affiliation{ARC Centre of Excellence for Engineered Quantum Systems, School of Mathematics and Physics, University of Queensland, Brisbane, QLD 4072, Australia.} 
\author{A. C. Doherty}
\affiliation{ARC Centre of Excellence for Engineered Quantum Systems, School of Physics, The University of Sydney, Sydney, NSW 2006, Australia.} 
\author{S. D. Barrett}
\affiliation{Blackett Laboratory and Institute for Mathematical Sciences, Imperial College London, London SW7 2PG, United Kingdom} 
\author{H. Lu}
\affiliation{Materials Department, University of California, Santa Barbara, California 93106, USA.}
\author{A. C. Gossard}
\affiliation{Materials Department, University of California, Santa Barbara, California 93106, USA.}
\author{D. J. Reilly$^\dagger$}
\affiliation{ARC Centre of Excellence for Engineered Quantum Systems, School of Physics, The University of Sydney, Sydney, NSW 2006, Australia.} 

\maketitle
{\bf 
The compound semiconductor gallium arsenide (GaAs) provides an ultra-clean platform \cite{Jesus_Nature_mat,Umansky} for storing and manipulating quantum information, encoded in the charge or spin states of electrons confined in nanostructures \cite{Hanson:2007eg}. The absence of inversion symmetry in the zinc-blende crystal structure of GaAs however, results in strong piezoelectric coupling between lattice acoustic phonons and electrons \cite{Fujisawa97,Brandes_Kramer}, a potential hinderance for quantum computing architectures that can be charge-sensitive during certain operations \cite{Hayashi,Petta_science05,Petta_PRL10,Yacoby2qubit,Granger_NatPhys}. Here we examine phonon generation in a GaAs double dot, configured as a single- or two-electron charge qubit, and driven by the application of microwaves via surface gates \cite{Petta_PRL04}. In a process that is a microwave analog of the Raman effect, stimulated phonon emission is shown to produce population inversion of a two-level system \cite{Dykman,Stace_PRL05} and provides spectroscopic signatures of the phononic environment created by the nanoscale device geometry.}

Devices based on GaAs are advantageous for hosting qubits because the electron's small effective mass in this material gives rise to a large level-splitting, the lack of valley degeneracy in the band structure simplifies operation, and the clean epitaxial interface used to confine electrons leads to inherently low charge noise \cite{Jesus_Nature_mat}. A potential drawback of GaAs and other group III-V compounds \cite{Nadj-Perge} is the presence of nuclear spins in the host lattice which can rapidly dephase electron spin-states \cite{Hanson:2007eg}. Dynamical-decoupling techniques\cite{MJB_DJR_NV} however, have recently addressed dephasing from nuclei, demonstrating \cite{Bluhm} that spin coherence can be preserved for times long enough that it is now important to address additional decoherence mechanisms such as residual charge noise and processes that incoherently couple electrons to phonons, either directly \cite{phonons}, or via the spin orbit interaction \cite{Stano06,Khaetskii00,Loss05}. In this respect, the piezoelectric nature of GaAs, while advantageous for shuttling electrons long distances\cite{Hermelin11,McNeil11}, also opens a channel for enhanced relaxation and dephasing, in particular, for qubit states with a charge dipole \cite{Petta_science05,Yacoby2qubit}.

Here we investigate a phonon emission process which, because of the strong piezoelectric coupling, limits charge coherence in GaAs, even in ideal structures at zero temperature.  Similar phonon generation mechanisms have recently been examined in the context of readout backaction \cite{Granger_NatPhys} and compared with transport measurements of InAs nanowires double dots \cite{Weber,Roulleau} and  graphene \cite{Roulleau}. Our test system is a charge qubit with one or two electrons in a double quantum dot, controlled by resonant microwaves\cite{Kouwenhoven_PRL94,Petta_PRL04} which drive Rabi oscillations of the electron between the ground and excited states, as shown schematically in Fig. 1(a). In the detuned regime where the microwave energy exceeds the qubit level splitting (see Fig. 1(b)), we show that this system undergoes stimulated phonon emission, a process which interrupts coherent driving and leads to population inversion, as predicted theoretically \cite{Dykman,Stace_PRL05}. This mechanism, which is a microwave version of the well known optical technique of Raman spectroscopy, provides a means of detecting the phonon spectral density created by the unique nanoscale device geometry. 
Our experimental results are in agreement with a theoretical model based on a non-Markovian master equation and we suggest approaches to suppress the electron-phonon coupling which could further improve coherence times.

\begin{figure*}
\includegraphics[scale=0.5]{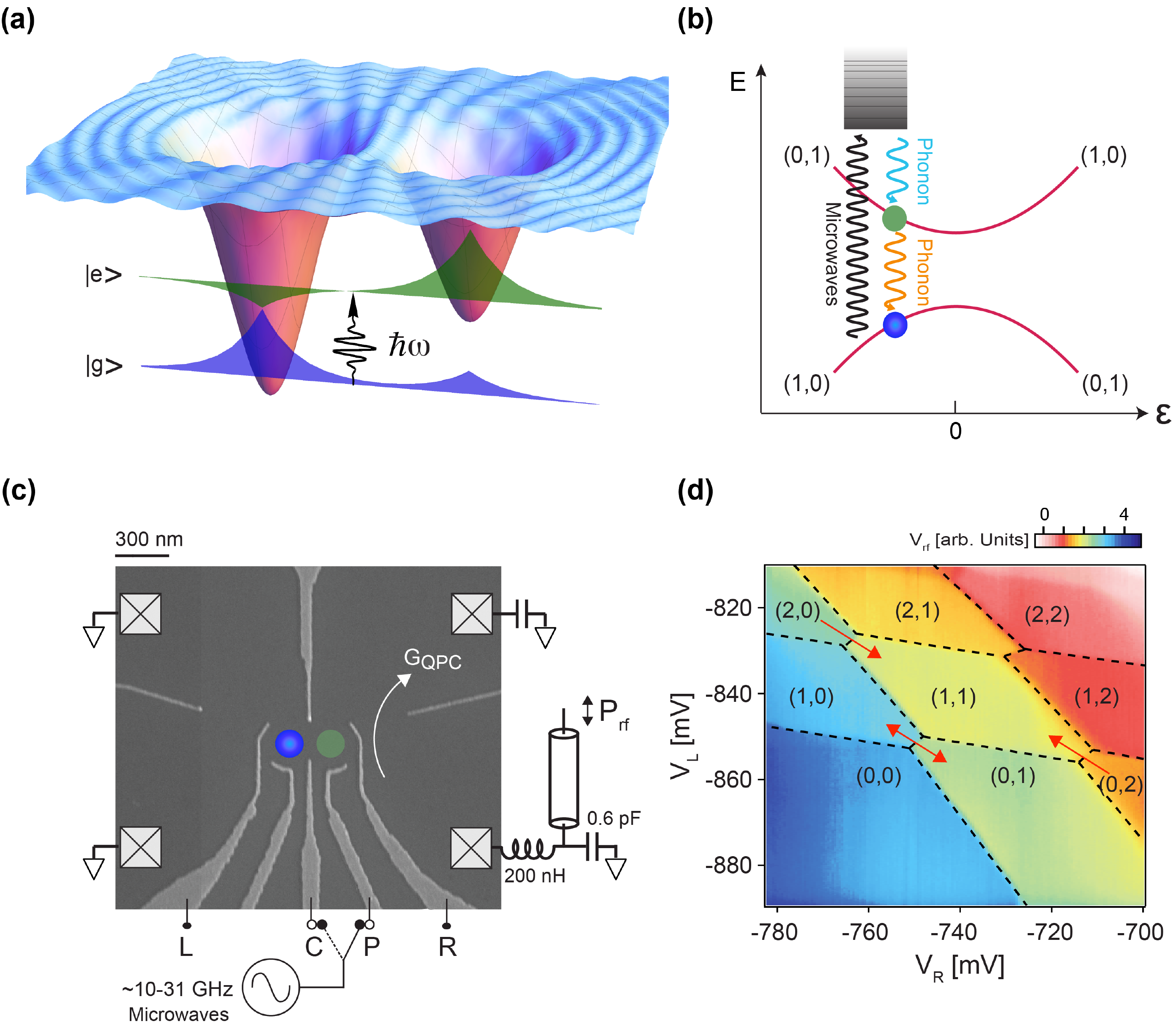}
\caption{\label{fig:epsart} {\bf (a)} Cartoon of the double dot potential showing a single electron wave function coherently tunnelling between the ground $|g\rangle$ and excited state $|e\rangle$ under microwave excitation. In a microwave analog of the Raman effect, stimulated emission of phonons (ripples) is modulated by the mode spectrum set by the intra-dot spacing, which for our device $\sim$ 280 nm. {\bf (b)} Energy-level diagram for the single-electron charge qubit showing the stimulated phonon emission process (light blue) that leads to asymmetric line shapes and population inversion. At a later time, spontaneous emission of a phonon (orange) leads to qubit relaxation. Grey shading depicts virtual states. {\bf (c)} Micrograph of the double dot device showing surface gates and ohmic contacts to the electron gas (crossed squares). Microwaves are applied to the plunger (P) or centre (C) gate. The conductance $G_{QPC}$ of a proximal rf-QPC detects the average charge state of the dot and modulates the amount of reflected rf power $P_{rf}$ from a resonant tank-circuit, enabling fast readout (see \S Methods for details).  {\bf (d)} Charge stability diagram of the double dot, detected using the rf-QPC. Labels (n,m) denote the number of electrons in the left and right quantum dots respectively. The demodulated signal $V_{rf}$ is proportional to the QPC conductance and thus the double dot charge configuration. Gate voltages $V_{L}$ and $V_{R}$ are applied to gates L and R in (c). Red arrows indicate the direction of allowed transitions under resonant microwave excitation. }
\end{figure*}

\begin{figure*}
\includegraphics[scale=0.5, trim = 0 0 0 50]{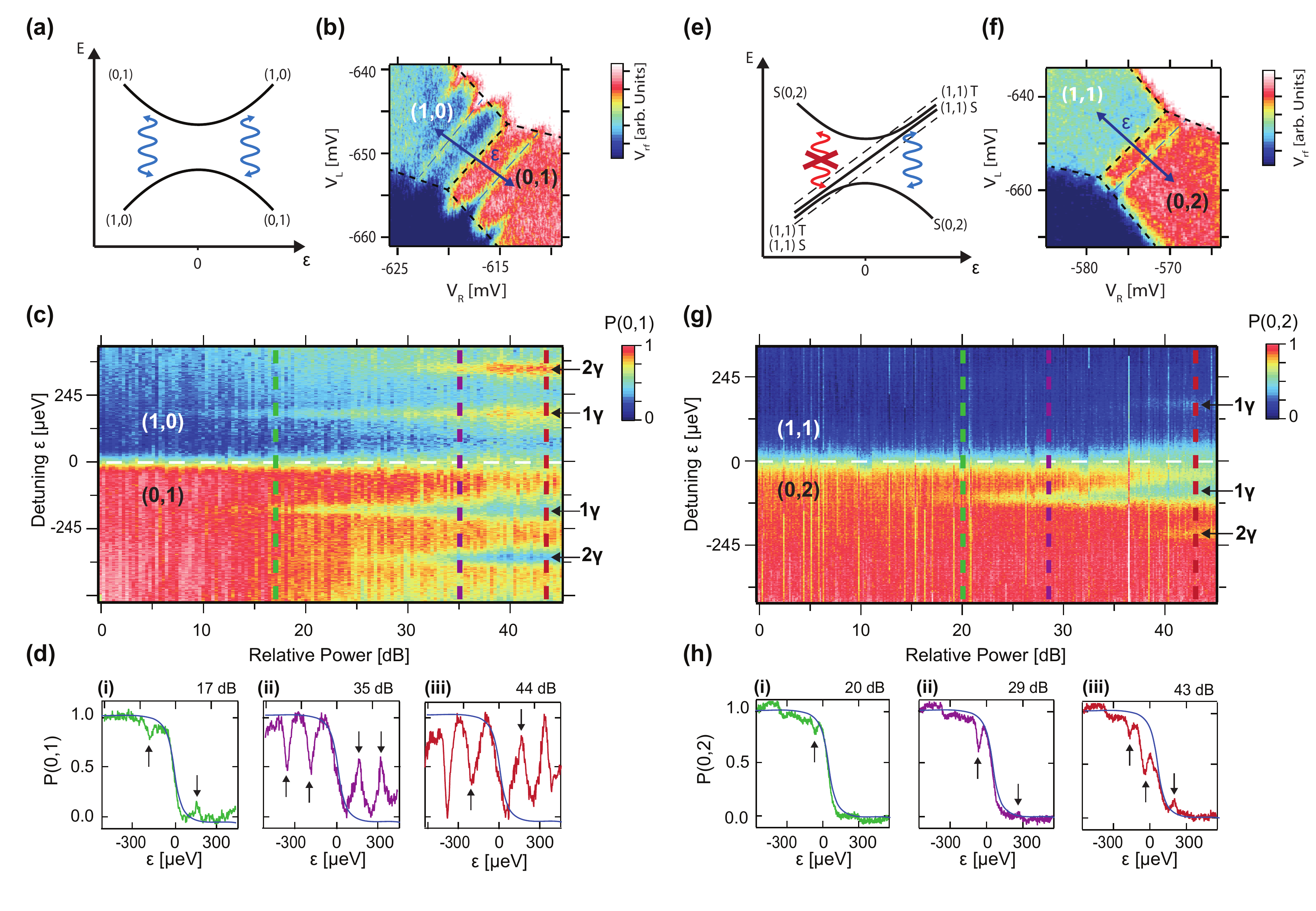}
\caption{\label{fig:epsart} {\bf (a)} Avoided crossing of the energy levels for the (0,1)-(1,0) transition under microwave excitation. Blue arrows indicate allowed microwave transitions. {\bf (b)} Charge stability diagram showing microwave sidebands either side of the (0,1)-(1,0) transition. {\bf (c)} Readout probability $P$(0,1) for an electron in the (0,1) state as a function of detuning $\epsilon$ and microwave power, where 0 dB is arbitrarily set to a power that yields no effect on the data. Microwave frequency $f$ = 31.8 GHz, applied to gate C. One- and two-photon sidebands (marked 1$\gamma$ and 2$\gamma$) are visible. {\bf (d)} Slices through (c) at different microwave powers, as indicated by the vertical dashed lines overlaying (c). {\bf (e)} Energy levels of the two-electron system under microwave excitation. Blue and red lines indicate different rates for microwave driving when the (0,2) singlet $S$ is the ground state, verse in (1,1) where the triplets $T$ are present. {\bf (f)} Stability diagram at the (0,2)-(1,1) transition with microwaves applied. Sidebands are visible in (0,2) but appear strongly suppressed in (1,1) due to Pauli spin-blockade. {\bf (g)} Readout probability $P$(0,2) for an electron in the (0,2) state as a function of detuning $\epsilon$ and microwave power. Microwave frequency is $f = 26.7 \text { GHz}$, applied to gate P. One- and two-photon sidebands (marked 1$\gamma$ and 2$\gamma$) are visible in (0,2) but are highly suppressed in the (1,1) regime. With increasing power these sidebands asymmetrically broaden on the blue-detuned side closest to $\epsilon$ = 0. {\bf (h)} Slices through (g) at positions indicated by the dashed lines in (g).}
\end{figure*}
\begin{figure*}
\includegraphics[scale=0.43, trim = 0 0 0 0]{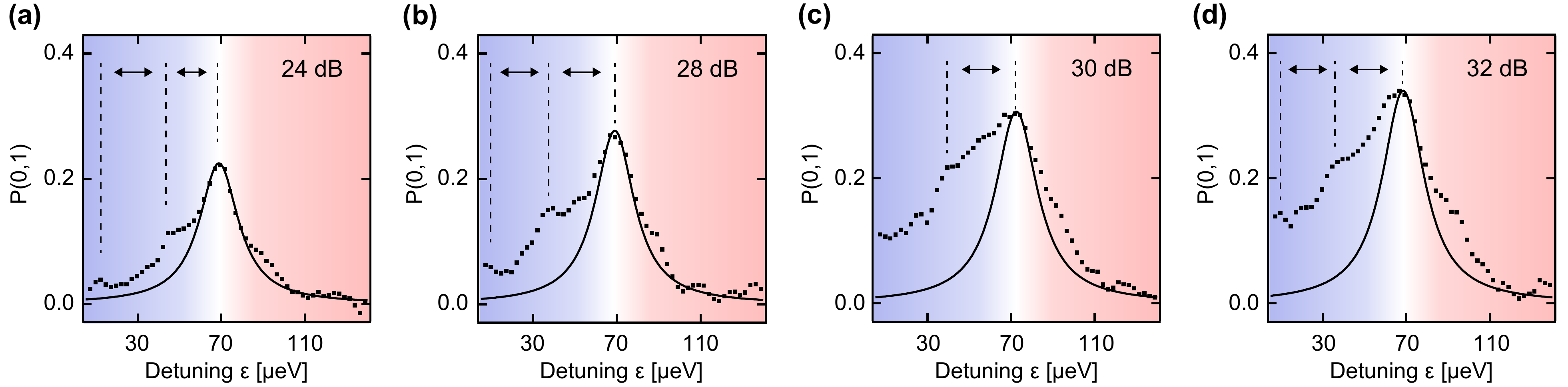}
\caption{\label{fig:epsart} {\bf (a)-(d)} Sideband lineshape for weak-driving in the one electron configuration, as a function of detuning and for increasing relative powers at $f$ = 31.8 GHz.  In the blue-detuned region ($\epsilon <\ \sim70\ \mu\text{eV}$), stimulated phonon emission results in asymmetric broadening of the sideband and the appearance of step-like features with a spacing (indicated by double arrows) that is set by the distance between the two quantum dots. Extracting this distance from the spacing of the steps gives 280 nm, consistent with the micrograph shown in Fig. 1(c). In the red-detuned region ($\varepsilon >\ \sim70\ \mu\text{eV}$), the sideband also deviates from the ideal Lorentzian shape, exhibiting additional skirting features which we attribute to a renormalization of the qubit levels due to coupling to the phononic environment. Solid lines are Lorentzian fits to the data using only data points in red region and at the top of the peak.}
\end{figure*}
\begin{figure}
\begin{center}
\includegraphics[scale=0.47, trim = 0 0 0 0]{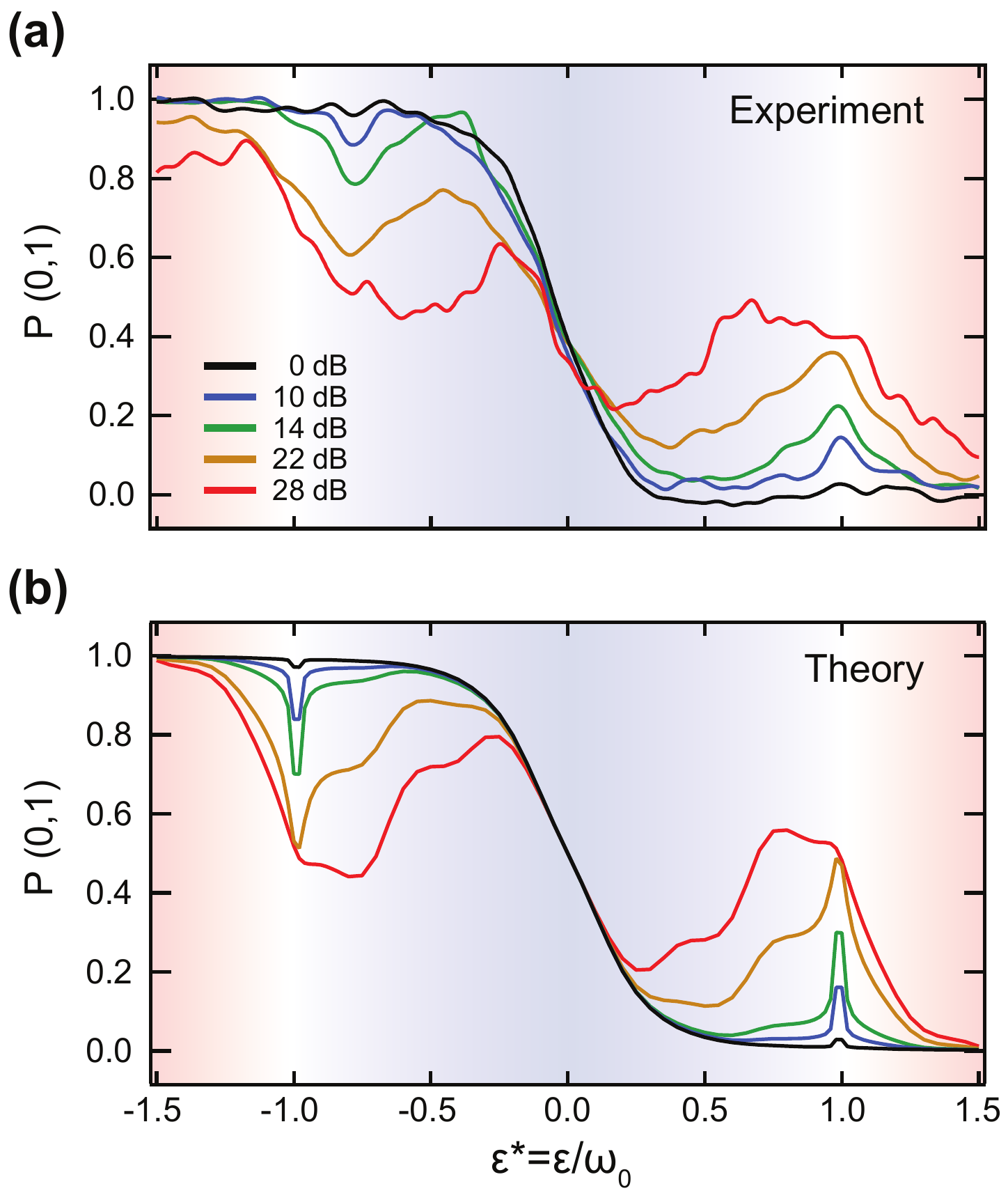}
\caption{{\bf (a)} Data showing the (0,1)-(1,0) transition and sidebands with microwaves applied to the C-gate. $\varepsilon^*$ is the dimensionless detuning, normalised by the microwave drive angular frequency $\omega_0$ = 2$\pi \times$ 32 GHz. {\bf (b)} Theoretical model using parameters consistent with experimental conditions (see \S Methods). Trace colours correspond to different microwave powers, consistent with data in (a). The left sideband in (a) occurs at $\varepsilon^*$ = -0.8, rather than -1, as should be expected.  This is likely an experimental artefact, possibly arising from a step-discontinuity when tuning the gate voltage near $\varepsilon^*$ = -0.2.}
\end{center}
\end{figure}
A micrograph of our double quantum dot device is shown in Fig. 1(c), including a proximal rf quantum point contact\cite{Reilly:2007ig} (rf-QPC) which is used as a sensor to read out the charge state of the system (see Fig. 1(d) and \S Methods). Gate voltages $V_L$ and $V_R$ control the detuning $\varepsilon$ of energy levels between the two dots. For $\varepsilon >>$ 0 the ground and excited states of the qubit correspond to localising the electron mostly in the left (1,0) or right (0,1) dot respectively, (this is reversed for $\varepsilon <<$ 0). We apply microwaves with an energy that matches the qubit splitting at a certain value of $\varepsilon$, coherently driving between ground and excited states. Under these conditions the readout signal exhibits sidebands that appear as lines in the charge stability diagram offset either side from the $\varepsilon$ = 0 transition (Fig. 2b)). We measure the time-averaged probability $P$ of the electron being in the (0,1) charge configuration,  calibrated such that $P$(0,1) = 1 for $\varepsilon\ll0$  and $P$(0,1) = 0 for $\varepsilon\gg0$. 

Close examination of the microwave sideband lineshape reveals that they are strongly asymmetric and distinct from the characteristic Lorentzian lineshape expected for a driven two-level system. This is seen clearly with increasing microwave power in Fig. 2(c), which shows pairs of sidebands corresponding to single- (1$\gamma$) and two- (2$\gamma$) microwave photon processes, positioned either side of the (1,0)-(0,1) transition. We note that the lineshape of all sidebands is strongly broadened, mostly on the side closest to $\varepsilon$ = 0, which we refer to as the blue-detuned side, where the microwave photon energy exceeds the qubit splitting. Further, at high powers, the amplitude of the microwave sidebands exceeds the saturation value of $P$(0,1) = 0.5 expected for a driven two-level system undergoing Rabi cycles between the ground and excited state. This population inversion results from Raman process in which the microwave energy exceeds the qubit level splitting, accessing the excited state by the stimulated emission of phonons at a rate that exceeds the rate of spontaneous decay \cite{Dykman,Stace_PRL05} (see Fig. 1(b)).

Our device can also be configured to the (2,0)-(1,1) charge transition by adjusting the gate potential to allow two electrons to occupy the double dot (see Fig. 1(d)). Under microwave excitation we again observe sidebands that are strongly asymmetric in their lineshape. A key difference in the two-electron case however, is the presence of Pauli-blockade \cite{Johnson_PRB05} which leads to spin dependent transitions when driving resonantly with microwaves \cite{Schreiber}. This behaviour is evident in Fig. 2(e-h) as a strong suppression in the sideband amplitude on the (1,1) side of the transition, where $\varepsilon >$ 0. We attribute this suppression to the occupation of a triplet state which cannot tunnel to the (0,2) singlet state under microwave excitation without a spin flip. The maximum height of the suppressed sideband in (1,1) is set by the ratio of singlets to triplets, (1:3 $\rightarrow$ (0.25 $\times$ $P$(0,2) = 0.5), gives $P$ = 0.125).  We find this spin-dependent suppression is unchanged for magnetic fields in the range $B$ = 0 -- 4 T \cite{foot1}. 

Even in the limit of weak microwave driving, where the system does not exhibit population inversion, stimulated emission of phonons will lead to decoherence of  charge states in GaAs devices. The probability for stimulated emission is weighted by the density of available phonon modes subject to the boundary conditions of the nanoscale device geometry. In an effort to uncover this geometric fingerprint in the lineshape, we make use of the high bandwidth of the rf-QPC charge detector to rapidly average over many data sets so that the sidebands can be observed with high resolution, as shown in Fig. 3(a-d) for a range of microwave powers in the weak driving regime. Comparing the averaged data to the Lorentzian lineshape expected for a weakly driven qubit in the Markovian regime \cite{StaceBarrett} (solid line), it is evident that  the blue-detuned region of the sideband shows fine, step-like features \cite{Stace_PRL05} in the excited state probability $P$(0,1) as a function of detuning. On the `red' side, the data also deviates slightly from the Lorentzian form and exhibits additional structure. Via comparison to a detailed theoretical model \cite{Stace_new}, we identify these aspects as signatures of the strong coupling between the driven qubit and its phononic environment.  

Our model describes the driven system with a master equation, allowing the Rabi frequency to be comparable to the decay rate. In order to incorporate Raman processes in the weak driving limit our model does not make the usual assumption of Markovian dynamics \cite{StaceBarrett,Stace_PRL05}. Taking the Laplace transform of the von Neumann equation gives a series of expressions that describe the dynamical steady state, dependent on the detuning $\epsilon$, the inter-dot tunnelling rate $\Delta$, the electric field-dipole Rabi frequency $\Omega$, the microwave driving frequency $\omega_0$, the temperature $T$, and the spectral density, $J(\omega)=2\pi\sum_q |g_q|^2\delta(\omega-\omega_q)$, where $g_q$ is the device geometry dependent electron-phonon coupling amplitude, $\omega_q$ is the phonon frequency and $\omega$ is the transition frequency. Figure 4 compares experimental data with our theoretical model using the materials properties of GaAs (see \S Methods for details). The only free parameter of this microscopic model is the global scaling of the microwave amplitude. The colour of the traces indicates the microwave driving amplitudes with values consistent between experiment (a) and theory (b). 

With this model in hand we comment further on several features that are evident in the experimental data and accounted for by the theory. Firstly, the step features appearing on the blue side of each sideband are understood in terms of a Raman transition involving the simultaneous absorption of a photon and emission of a phonon, weighted by the phonon spectral density $J(\omega)$ and set by the device geometry. The width of each step (see Fig. 3) is directly related to the spacing between the quantum dots and can be considered as a double-slit interference pattern for emitting a phonon on the left or right dots during microwave drive.  A shoulder to the lineshape on the red-side of the sideband is also evident and arises from a renormalisation of the qubit detuning and Rabi frequency when the bare electron interacts with the phononic environment of the crystal lattice. This renormalisation of the detuning is analogous to the Lamb-shift, and arises in Markovian models, whilst the renormalised Rabi frequency only appears when accounting for the dynamical steady-state of the driven system.

There are also some discrepancies between our model and the experimental data. These can likely be explained by the anisotropy of the piezoelectric coupling, which we have neglected in the model. Further, we have not considered the presence of the surface which modifies the phononic spectral density. For the present device, where the double dots are located 110 nm below the surface, constructive interference between the double-dot dipole and its image charge couple the electrons to Rayleigh surface acoustic waves. We note that the ability to control the crystallographic orientation of the double dot and its depth from the surface offers a means of suppressing electron-phonon coupling, an advantage of heterostructure devices. Future approaches to suppressing the influence of the phononic environment may include patterning the surface or shaping the gate electrodes to induce phononic band gaps \cite{Painter} that extend qubit coherence in these systems.   

\begin{section}{Methods}
The double dot is defined electrostatically, 110 nm below the surface of a GaAs/Al$_{0.3}$Ga$_{0.7}$As heterostructure grown using molecular beam epitaxy (electron density 2.4 $\times$ 10 $^{15}$ m $^{-2}$,
mobility 44 m $^{2}$/Vs at 20 K). All data is taken at the base electron temperature of a dilution refrigerator, $T_e\sim$ 100 mK, with the sample mounted on a custom high frequency printed circuit board (PCB) \cite{CollessRSI}. Microwave excitation is produced using a room temperature vector source (Agilent 8267D) and fed to the device PCB via coaxial cables that include cryogenic attenuators. 

Readout is performed using an rf quantum point contact (rf-QPC), proximal to the double dot. An impedance matching tank circuit operating at a frequency of $\sim$ 500 MHz transforms the high QPC resistance towards the 50 $\Omega$ characteristic impedance of a transmission line enabling the QPC to modulate the amount of reflected rf power. The change in reflected rf power is amplified using cryogenic and room temperature amplifiers and demodulated at room temperature using standard quadrature mixing techniques to yield a baseband signal $V_{rf}$ proportional to the QPC conductance. For high resolution data (Figure 3 and Figure 4) a high bandwidth digital storage scope is used to perform a large number of trace averages. 

To compare our theoretical model to the experimental data, we normalise each quantity with respect to the microwave driving frequency, e.g.,  $\omega^*=\omega/\omega_0$, where the $^*$ indicates dimensionless parameters. In this form we can write: 
\begin{equation}
J^*(\omega^*)=\pi\mathsf{P^*}\omega^*\frac{1-\mathrm{sinc}(d^* \omega^*)}{1+(\omega^*/\omega_c^*)^2},\label{eqn:J}
\end{equation}
where $d^*=d\,\omega_0/c_s$, with $d$ the inter-dot separation and $c_s$ is the transverse speed of sound. $\mathsf{P^*}=(\hbar P)^2/(4\pi^2\hbar \mu\,c_s^3)$ where $\hbar P$ is the piezoelectric electron-phonon coupling strength, $\mu$ is the mass density, and $\omega_c\approx2\pi c_s/a$ is a high-frequency cut-off determined by the exponential decay length of the localised electronic wavefunction, $a$.  
For GaAs, $c_s=3000$ m/s,  $\hbar P=1.45$ eV/nm and $\mu=5300\textrm{ kg/m}^3$, so  $P^*=0.09$. For the driving frequency $\omega_0=2\pi\times 32$ GHz pertinent to Fig. 4(a), and assuming $d\sim 280$ nm (set by device geometry) we find $d^*\sim 20$.  The tunnelling rate $\Delta^*=0.15$, and temperature $T^*=k_BT/\omega_0=0.12$ are obtained independently from experimental data. We choose  $\omega_c^*=2$, consistent with $a \sim 50$ nm. 
\end{section}

\vspace{0.5cm}
$\dagger$ Corresponding author, email: david.reilly@sydney.edu.au \\

$*$ These authors contributed equally to this work.

\begin{section}{Acknowledgements}
We acknowledge funding from the U.S. Intelligence Advanced Research Projects Activity (IARPA), through the U.S. Army Research Office and the Australian Research Council Centre of Excellence Scheme (Grant No. EQuS CE110001013).\\
\end{section}
\small
\bibliographystyle{naturemag}

\begin{thebibliography}{10}
\expandafter\ifx\csname url\endcsname\relax
  \def\url#1{\texttt{#1}}\fi
\expandafter\ifx\csname urlprefix\endcsname\relax\def\urlprefix{URL }\fi
\providecommand{\bibinfo}[2]{#2}
\providecommand{\eprint}[2][]{\url{#2}}

\bibitem{Jesus_Nature_mat}
\bibinfo{author}{del Alamo, J.~A.}
\newblock \bibinfo{title}{{Nanometre-scale electronics with III-V compound
  semiconductors}}.
\newblock \emph{\bibinfo{journal}{Nature}} \textbf{\bibinfo{volume}{479}},
  \bibinfo{pages}{317--323} (\bibinfo{year}{2011}).

\bibitem{Umansky}
\bibinfo{author}{Umansky, V.} \emph{et~al.}
\newblock \bibinfo{title}{{MBE growth of ultra-low disorder 2DEG with mobility
  exceeding $35\times10^6$ cm$^2$/V s}}.
\newblock \emph{\bibinfo{journal}{J. Crystal Growth}}
  \textbf{\bibinfo{volume}{311}}, \bibinfo{pages}{1658--1661}
  (\bibinfo{year}{2009}).

\bibitem{Hanson:2007eg}
\bibinfo{author}{Hanson, R.}, \bibinfo{author}{Petta, J.~R.},
  \bibinfo{author}{Tarucha, S.} \& \bibinfo{author}{Vandersypen, L. M.~K.}
\newblock \bibinfo{title}{{Spins in few-electron quantum dots}}.
\newblock \emph{\bibinfo{journal}{Rev. Mod. Phys.}}
  \textbf{\bibinfo{volume}{79}}, \bibinfo{pages}{1217--1265}
  (\bibinfo{year}{2007}).

\bibitem{Fujisawa97}
\bibinfo{author}{Fujisawa, T.} \emph{et~al.}
\newblock \bibinfo{title}{Spontaneous emission spectrum in double quantum dot
  devices}.
\newblock \emph{\bibinfo{journal}{Science}} \textbf{\bibinfo{volume}{282}},
  \bibinfo{pages}{932--935} (\bibinfo{year}{1998}).

\bibitem{Brandes_Kramer}
\bibinfo{author}{Brandes, T.} \& \bibinfo{author}{Kramer, B.}
\newblock \bibinfo{title}{Spontaneous emission of phonons by coupled quantum
  dots}.
\newblock \emph{\bibinfo{journal}{Phys. Rev. Lett.}}
  \textbf{\bibinfo{volume}{83}}, \bibinfo{pages}{3021--3024}
  (\bibinfo{year}{1999}).

\bibitem{Hayashi}
\bibinfo{author}{Hayashi, T.}, \bibinfo{author}{Fujisawa, T.},
  \bibinfo{author}{Cheong, H.}, \bibinfo{author}{Jeong, Y.~H.} \&
  \bibinfo{author}{Hirayama, Y.}
\newblock \bibinfo{title}{Coherent manipulation of electronic states in a
  double quantum dot}.
\newblock \emph{\bibinfo{journal}{Phys. Rev. Lett.}}
  \textbf{\bibinfo{volume}{91}}, \bibinfo{pages}{226804}
  (\bibinfo{year}{2003}).

\bibitem{Petta_science05}
\bibinfo{author}{Petta, J.~R.} \emph{et~al.}
\newblock \bibinfo{title}{{Coherent manipulation of coupled electron spins in
  semiconductor quantum dots}}.
\newblock \emph{\bibinfo{journal}{Science}} \textbf{\bibinfo{volume}{309}},
  \bibinfo{pages}{2180--2184} (\bibinfo{year}{2005}).

\bibitem{Petta_PRL10}
\bibinfo{author}{Petersson, K.~D.}, \bibinfo{author}{Petta, J.~R.},
  \bibinfo{author}{Lu, H.} \& \bibinfo{author}{Gossard, A.~C.}
\newblock \bibinfo{title}{Quantum coherence in a one-electron semiconductor
  charge qubit}.
\newblock \emph{\bibinfo{journal}{Phys. Rev. Lett.}}
  \textbf{\bibinfo{volume}{105}}, \bibinfo{pages}{246804}
  (\bibinfo{year}{2010}).

\bibitem{Yacoby2qubit}
\bibinfo{author}{Shulman, M.~D.} \emph{et~al.}
\newblock \bibinfo{title}{Demonstration of entanglement of electrostatically
  coupled single-triplet qubits}.
\newblock \emph{\bibinfo{journal}{Science}} \textbf{\bibinfo{volume}{336}},
  \bibinfo{pages}{202--205} (\bibinfo{year}{2012}).

\bibitem{Granger_NatPhys}
\bibinfo{author}{Granger, G.} \emph{et~al.}
\newblock \bibinfo{title}{Quantum interference and phonon-mediated back-action
  in lateral quantum-dot circuits}.
\newblock \emph{\bibinfo{journal}{Nature Phys.}} \textbf{\bibinfo{volume}{8}},
  \bibinfo{pages}{522--527} (\bibinfo{year}{2012}).

\bibitem{Petta_PRL04}
\bibinfo{author}{Petta, J.~R.}, \bibinfo{author}{Johnson, A.~C.},
  \bibinfo{author}{Marcus, C.~M.}, \bibinfo{author}{Hanson, M.~P.} \&
  \bibinfo{author}{Gossard, A.~C.}
\newblock \bibinfo{title}{Manipulation of a single charge in a double quantum
  dot}.
\newblock \emph{\bibinfo{journal}{Phys. Rev. Lett}}
  \textbf{\bibinfo{volume}{93}}, \bibinfo{pages}{186802}
  (\bibinfo{year}{2004}).

\bibitem{Dykman}
\bibinfo{author}{Dykman, M.~I.}
\newblock \bibinfo{title}{Relaxation of impurities in a nonresonant field and
  phonon amplification}.
\newblock \emph{\bibinfo{journal}{Sov. J. Low Temp. Phys.}}
  \textbf{\bibinfo{volume}{5}}, \bibinfo{pages}{89--95} (\bibinfo{year}{1979}).

\bibitem{Stace_PRL05}
\bibinfo{author}{Stace, T.~M.}, \bibinfo{author}{Barrett, S.~D.} \&
  \bibinfo{author}{Doherty, A.~C.}
\newblock \bibinfo{title}{Population inversion of a driven two-level system in
  a structureless bath}.
\newblock \emph{\bibinfo{journal}{Phys. Rev. Lett.}}
  \textbf{\bibinfo{volume}{95}}, \bibinfo{pages}{106801}
  (\bibinfo{year}{2005}).

\bibitem{Nadj-Perge}
\bibinfo{author}{Nadj-Perge, S.}, \bibinfo{author}{Frolov, S.~M.},
  \bibinfo{author}{Bakkers, E. P. A.~M.} \& \bibinfo{author}{Kouwenhoven,
  L.~P.}
\newblock \bibinfo{title}{Spin-orbit qubit in a semiconductor nanowire}.
\newblock \emph{\bibinfo{journal}{Nature}} \textbf{\bibinfo{volume}{468}},
  \bibinfo{pages}{1084--1087} (\bibinfo{year}{2010}).

\bibitem{MJB_DJR_NV}
\bibinfo{author}{Biercuk, M.~J.} \& \bibinfo{author}{Reilly, D.~J.}
\newblock \bibinfo{title}{Solid-state spins survive}.
\newblock \emph{\bibinfo{journal}{Nature Nanotechnology}}
  \textbf{\bibinfo{volume}{6}}, \bibinfo{pages}{9} (\bibinfo{year}{2011}).

\bibitem{Bluhm}
\bibinfo{author}{Bluhm, H.} \emph{et~al.}
\newblock \bibinfo{title}{Dephasing time of gaas electron-spin qubits coupled
  to a nuclear bath exceeding 200 us}.
\newblock \emph{\bibinfo{journal}{Nature Phys.}} \textbf{\bibinfo{volume}{7}},
  \bibinfo{pages}{109--113} (\bibinfo{year}{2010}).

\bibitem{phonons}
\bibinfo{author}{Fedichkin, L.} \& \bibinfo{author}{Fedorov, A.}
\newblock \bibinfo{title}{Decoherence rate of a semiconductor charge quit
  coupled to acoustic phonon reservoir}.
\newblock \emph{\bibinfo{journal}{Phys. Rev. A.}}
  \textbf{\bibinfo{volume}{69}}, \bibinfo{pages}{032311}
  (\bibinfo{year}{2004}).

\bibitem{Stano06}
\bibinfo{author}{Stano, P.} \& \bibinfo{author}{Fabian, J.}
\newblock \bibinfo{title}{Theory of phonon-induced spin relaxation in laterally
  coupled quantum dots}.
\newblock \emph{\bibinfo{journal}{Phys. Rev. Lett.}}
  \textbf{\bibinfo{volume}{96}}, \bibinfo{pages}{186602}
  (\bibinfo{year}{2006}).

\bibitem{Khaetskii00}
\bibinfo{author}{Khaetskii, A.~V.} \& \bibinfo{author}{Nazarov, Y.~V.}
\newblock \bibinfo{title}{Spin relaxation in semiconductor quantum dots}.
\newblock \emph{\bibinfo{journal}{Phys. Rev. B}} \textbf{\bibinfo{volume}{61}},
  \bibinfo{pages}{12639} (\bibinfo{year}{2000}).

\bibitem{Loss05}
\bibinfo{author}{Bulaev, D.~V.} \& \bibinfo{author}{Loss, D.}
\newblock \bibinfo{title}{Spin relaxation and anticrossing in quantum dots:
  Rashba versus dresselhaus spin-orbit coupling}.
\newblock \emph{\bibinfo{journal}{Phys. Rev. B}} \textbf{\bibinfo{volume}{71}},
  \bibinfo{pages}{205324} (\bibinfo{year}{2005}).

\bibitem{Hermelin11}
\bibinfo{author}{Hermelin, S.} \emph{et~al.}
\newblock \bibinfo{title}{Electrons surfing on a sound wave as a platform for
  quantum optics with flying electrons}.
\newblock \emph{\bibinfo{journal}{Nature}} \textbf{\bibinfo{volume}{477}},
  \bibinfo{pages}{435--438} (\bibinfo{year}{2011}).

\bibitem{McNeil11}
\bibinfo{author}{McNeil, R. P.~G.} \emph{et~al.}
\newblock \bibinfo{title}{On-demand single-electron transfer between distant
  quantum dots}.
\newblock \emph{\bibinfo{journal}{Nature}} \textbf{\bibinfo{volume}{477}},
  \bibinfo{pages}{439--442} (\bibinfo{year}{2011}).

\bibitem{Weber}
\bibinfo{author}{Weber, C.} \emph{et~al.}
\newblock \bibinfo{title}{{Probing Confined Phonon Modes by Transport through a
  Nanowire Double Quantum Dot}}.
\newblock \emph{\bibinfo{journal}{Phys. Rev. Lett.}}
  \textbf{\bibinfo{volume}{104}}, \bibinfo{pages}{036801}
  (\bibinfo{year}{2010}).

\bibitem{Roulleau}
\bibinfo{author}{Roulleau, P.} \emph{et~al.}
\newblock \bibinfo{title}{Coherent electron-phonon coupling in tailored quantum
  systems}.
\newblock \emph{\bibinfo{journal}{Nature Com.}} \textbf{\bibinfo{volume}{2}},
  \bibinfo{pages}{239} (\bibinfo{year}{2011}).

\bibitem{Kouwenhoven_PRL94}
\bibinfo{author}{Kouwenhoven, L.~P.} \emph{et~al.}
\newblock \bibinfo{title}{Observation of photon-assisted tunnelling through a
  quantum dot}.
\newblock \emph{\bibinfo{journal}{Phys. Rev. Lett}}
  \textbf{\bibinfo{volume}{73}}, \bibinfo{pages}{3443--2446}
  (\bibinfo{year}{1994}).

\bibitem{Reilly:2007ig}
\bibinfo{author}{Reilly, D.~J.}, \bibinfo{author}{Marcus, C.~M.},
  \bibinfo{author}{Hanson, M.~P.} \& \bibinfo{author}{Gossard, A.~C.}
\newblock \bibinfo{title}{{Fast single-charge sensing with a rf quantum point
  contact}}.
\newblock \emph{\bibinfo{journal}{App. Phys. Lett.}}
  \textbf{\bibinfo{volume}{91}}, \bibinfo{pages}{162101}
  (\bibinfo{year}{2007}).

\bibitem{Johnson_PRB05}
\bibinfo{author}{Johnson, A.~C.} \emph{et~al.}
\newblock \bibinfo{title}{Singlet-triplet spin blockade and charge sensing in a
  few-electron double quantum dot}.
\newblock \emph{\bibinfo{journal}{Phys. Rev. B.}}
  \textbf{\bibinfo{volume}{72}}, \bibinfo{pages}{165308}
  (\bibinfo{year}{2005}).

\bibitem{Schreiber}
\bibinfo{author}{Schreiber, L.~R.} \emph{et~al.}
\newblock \bibinfo{title}{{Coupling artificial molecular spin states by
  photon-assisted tunnelling}}.
\newblock \emph{\bibinfo{journal}{Nature Com.}} \textbf{\bibinfo{volume}{2}},
  \bibinfo{pages}{556} (\bibinfo{year}{2011}).

\bibitem{foot1}
\bibinfo{title}{This dependence is somewhat in contrast to the recent work by
  schreiber $et$ $al.,$ \cite{Schreiber} where spin blockade is lifted with
  microwave excitation, perhaps due to the presence of a micro-magnet on the
  surface of their device}.

\bibitem{StaceBarrett}
\bibinfo{author}{Barrett, S.~D.} \& \bibinfo{author}{Stace, T.}
\newblock \bibinfo{title}{Continuous measurement of a microwave-driven solid
  state qubit}.
\newblock \emph{\bibinfo{journal}{Phys. Rev. Lett}}
  \textbf{\bibinfo{volume}{96}}, \bibinfo{pages}{017405}
  (\bibinfo{year}{2006}).

\bibitem{Stace_new}
\bibinfo{author}{Stace, T.~M.}, \bibinfo{author}{Doherty, A.~C.} \&
  \bibinfo{author}{Reilly, D.~J.}
\newblock \bibinfo{title}{In review}.

\bibitem{Painter}
\bibinfo{author}{Alegre, T. P.~M.}, \bibinfo{author}{Safavi-Naeini, A.},
  \bibinfo{author}{Winger, M.} \& \bibinfo{author}{Painter, O.}
\newblock \bibinfo{title}{Quasi-two-dimensional optomechanical crystals with a
  complete phononic bandgap}.
\newblock \emph{\bibinfo{journal}{Optics Express}}
  \textbf{\bibinfo{volume}{19}}, \bibinfo{pages}{5658} (\bibinfo{year}{2011}).

\bibitem{CollessRSI}
\bibinfo{author}{Colless, J.~I.} \& \bibinfo{author}{Reilly, D.~J.}
\newblock \bibinfo{title}{Cryogenic high-frequency readout and control platform
  for spin qubits}.
\newblock \emph{\bibinfo{journal}{Rev. Sci. Instrum.}}
  \textbf{\bibinfo{volume}{83}}, \bibinfo{pages}{023902}
  (\bibinfo{year}{2012}).

\end{thebibliography}

\end{document}